\documentclass[12pt]{article}

\newcommand{\prl}[3]{{\it Phys. Rev. Lett.} {\bf #1},  #2 (#3)}
\newcommand{\prd}[3]{{\it Phys. Rev.} D {\bf #1},  #2 (#3)}
\newcommand{\plb}[3]{{\it Phys. Lett.} B {\bf #1},  #2 (#3)}

\newcommand{\npb}[3]{{\it Nucl. Phys.} B {\bf #1},  #2 (#3)}

\newcommand{\epc}[3]{{\it Eur. Phys.} J. C {\bf #1},  #2 (#3)}

\newcommand{\pr}[3]{{\it Phys. Rep.} {\bf #1},  #2 (#3)}

\def\etal{{\it et al.}}

\def\Gevc{GeV/$c$}
\def\Gevcsq{GeV/$c^2$}

\def\Mevcsq{MeV/$c^2$}

\def\ebeam{E_{\rm beam}^*}
\def\dE{\Delta E}
\def\DE{$\dE$}
\def\mb{M_{\rm bc}}
\def\Mb{$\mb$}
\def\micron{$\mu$m}

\def\BF{{\cal B}}

\def\vcb{V_{cb}}
\def\vub{V_{ub}}
\def\thrust{\theta_{\rm T}}

\def\bzb{{\overline{B}{}^0}}
\def\bz{{B^0}}
\def\bm{{B^-}}
\def\bplus{{B^+}}

\def\Fb{fb$^{-1}$}
\def\piz{\pi^0}
\def\pip{\pi^+}
\def\pim{\pi^-}
\def\kz{K^0}
\def\kp{K^+}
\def\km{K^-}
\def\ks{K^0_S}

\def\bbar{\overline{B}}
\def\bbbar{B\bbar}

\def\BBbar{$\bbbar$}

\def\ccbar{c\overline{c}}

\def\dcp{D_{CP}}

\def\dstarp{D^{*+}}
\def\dstp{D^{*+}}
\def\dstm{D^{*-}}
\def\dstz{D^{*0}}

\def\nub{\overline{\nu}}

\def\dz{D^0}

\def\dminus{D^-}
\def\dplus{D^+}

\def\dzb{\overline{D}{}^0}

\def\bzdstlnu{\bzb\to\dstarp\ell^-\nub}

\def\kpi{\km\pi^+}
\def\kpipiz{\km\pip\piz}
\def\kpipipi{\km\pip\pip\pim}
\def\kbar{\overline{K}}
\def\kzb{\overline{K}{}^{0}}
\def\kstb{\overline{K}{}^{*}}
\def\kstzb{\overline{K}{}^{*0}}
\def\kstm{{K}{}^{*-}}

\def\Ups{\Upsilon(4\mathrm{S})}
\def\UPS{$\Ups$}
\def\fz{f_0}

\usepackage{graphicx}

\begin{document}



\resizebox{0.2\textwidth}{!}{\includegraphics{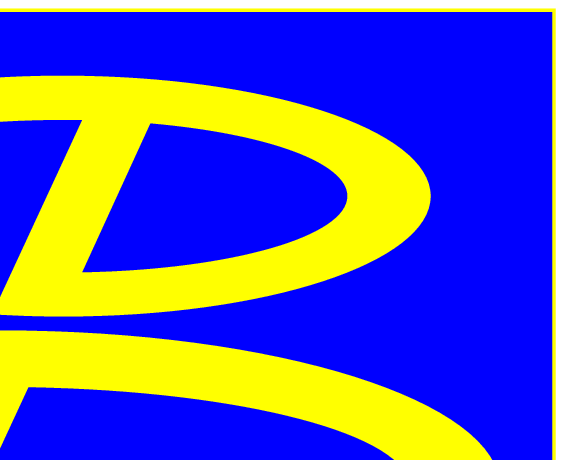}}

\vskip -3cm
\noindent
\hspace*{4.5in}BELLE Preprint 2001-16 \\
\hspace*{4.5in}KEK Preprint 2001-136 \\
\begin{center}
\vskip 2cm
{\Large  \bf
Belle $B$ Physics Results}

\vskip 1cm

{\large H. Tajima\\
(The Belle Collaboration)\\
\smallskip
Department of Physics, University of Tokyo, \\
7-3-1 Hongo, Tokyo 113-0033 Japan\\
 E-mail: tajima@phys.s.u-tokyo.ac.jp}\\
 
\vskip 1cm

\end{center}

\smallskip

\begin{abstract}
$B$ physics results from the Belle experiment are reviewed. 
Precise measurements of Cabbibo-Kobayashi-Maskawa matrix elements are made.
Several decay modes are observed which will enable us to measure the $CP$ angles, $\phi_i$ in channels other than $\bz\to\psi\kz$ modes.
New rare decay modes are observed in many channels.
Particular attention is paid to the first observation of the electroweak penguin-mediated decay.
\end{abstract}

\bigskip
\bigskip

\begin{center}
Contributed to the Proceedings of the XX International Symposium \\
on Lepton and Photon Interactions at High Energies, \\
July 23--28, 2001, Rome, Italy.
\end{center}
\thispagestyle{empty}

\newpage

\section{Introduction}
One of the main objectives of heavy flavor physics is the
determination of 
the Cabbibo-Kobayashi-Maskawa (CKM) quark mixing matrix.\cite{CKM}
In the hadronic charged current of weak decay, 
the weak and flavor eigenstates are not identical. 
The CKM matrix describes the relation between these two sets of states.
An irreducible phase in this matrix, first introduced
by Kobayashi and Maskawa, gives rise to $CP$ violation 
in the framework of the Standard Model (SM).

The unitarity of the CKM matrix for $b$ and $d$ quark sectors leads to the expression
\begin{eqnarray}
V_{ud}V_{ub}^* + V_{cd}V_{cb}^* + V_{td}V_{tb}^* = 0.
\end{eqnarray}
This relation can be displayed as a triangle in the complex plane as shown in \begin{figure}[h]
\center{\resizebox{0.50\textwidth}{!}{\includegraphics{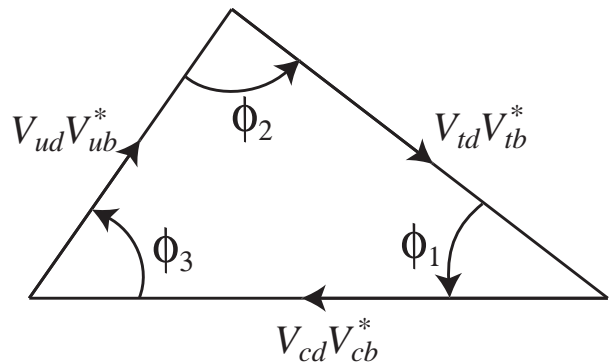}}}
\caption{Unitarity triangle.}
\label{fig:CKM-triangle}
\end{figure}
Figure~\ref{fig:CKM-triangle}.
The angles of the triangle are defined as\cite{phi}
\begin{eqnarray}
\phi_1 &\equiv& \pi - \arg\left(\frac{-V_{td}V_{tb}^*}{-V_{cd}V_{cb}^*}\right)=\beta,\nonumber\\
\phi_2 &\equiv& \arg\left(\frac{V_{td}V_{tb}^*}{-V_{ud}V_{ub}^*}\right)=\gamma,\label{eq:phi}\\
\phi_3 &\equiv& \arg\left(\frac{V_{ud}V_{ub}^*}{-V_{cd}V_{cb}^*}\right)=\alpha.\nonumber
\end{eqnarray}
A crucial test of the SM is to evaluate the consistency of all sides
and angles of this unitarity triangle. This test
 may provide information on the dynamical origin of the quark mixing matrix.

Experimentally, the magnitudes of the elements of the
CKM matrix can be determined by measuring decay rates 
(including the mixing rate) and production rates.
The angles $\phi_i$ can be determined by measurements of $CP$ 
asymmetries caused by the interference of two or more amplitudes 
with different weak phases.
For example, $\sin2\phi_1$ can be determined by the $CP$ asymmetry induced by interference between mixing which carries the $V_{tb}^*$ phase and $\bz\to\psi\kz$  decay which carries the $V_{cb}^*$ phase as indicated by Eq.~\ref{eq:phi}.

In this paper, we review the $B$ physics results (except for the results which involve decay-time dependent analysis) from the Belle experiment.
We report the determination of the magnitude of 
the CKM matrix element $|\vcb|$ and the observation of 
new $B$ decay modes which will lead to measurements of 
the $CP$ angles $\phi_i$.

\section{The Belle Detector}
The Belle detector is designed and constructed primarily to observe and measure $CP$ violation in $B$ decays.
Because of the high luminosity of KEKB\cite{KEKB} ($L_{peak}\approx 5.0\times10^{33} \mathrm{cm}^{-2}\mathrm{s}^{-1}$), Belle currently accumulates data at a rate of more than 4.5~\Fb\ per month.
This corresponds to 5 million \BBbar\ events per month,
allowing precise measurements of $B$ and charmed hadron properties.
Most of the analyses presented in this paper are based on an integrated luminosity of 21.3~\Fb.

Due to the asymmetric energies of the colliding beams (3.5 GeV $\times$ 8 GeV), the \UPS\ and its daughter $B$ mesons are produced at $\beta\gamma \approx 0.425$ in the laboratory frame: the difference in $B$ meson decay times may be measured using the difference in decay vertex positions. 
This key feature of the Belle experiment allows the measurement of $CP$ violation through mixing in neutral $B$ decays.

The Belle detector consists of a silicon vertex detector (SVD), a central drift chamber (CDC), an aerogel $\check{\mathrm{C}}$erenkov counter (ACC), a time of flight (TOF) and trigger scintillation counter (TSC) system, an electromagnetic calorimeter (ECL), and a $K_L$/muon detector (KLM).

The SVD measures the precise position of decay vertices.
It consists of three layers of double-sided silicon strip detectors (DSSD) in a barrel-only design and covers 86\% of solid angle.
The three layers are at radii of 3.0, 4.5 and 6 cm respectively.
Impact parameter resolutions are measured as functions of momentum $p$ (\Gevc) to be $\sigma_{xy}^2=19^2 + [50/(p\beta\sin^{3/2}\theta)]^2$ \micron${}^2$ and $\sigma_{z}^2=36^2 + [42/(p\beta\sin^{5/2}\theta)]^2$ \micron${}^2$, where $\theta$ is the polar angle with respect to the beam direction.

Charged tracks are primarily recognized by the CDC.
The CDC covers 
92\% of solid angle in the center of mass (CM) frame, and  
consists of 50 cylindrical layers of drift cells organized into 11 super-layers each containing between three and six layers. 
He-C${}_2$H${}_6$ (50/50\%) gas is used to minimize multiple-Coulomb scattering. 
The magnetic field of 1.5~Tesla is chosen to optimize momentum resolution without sacrificing efficiency for low momentum tracks. 
The transverse momentum resolution for charged tracks is
$(\sigma_{p_T}/p_T)^2=(0.0019 p_T)^2+(0.0030/\beta)^2$, where $p_T$ is in GeV/$c$.

Particle identification is accomplished by a combination of the ACC, the TOF and the CDC. 
The combination of these particle identification devices is a key feature of the Belle detector.
The combined response of the three systems provide $K^\pm$ identification with an efficiency of about 85\% and a charged pion fake rate of about 10\% for all momenta up to 3.5~GeV/$c$.

The CDC and ECL are used to identify electrons. 
The ECL also detects photons and measures their energy. 
The ECL consists of 30~cm (16.1$X_0$) long Cesium Iodide (Tl) crystals. 
The photon energy resolution is $(\sigma_E/E)^2=(0.013)^2+(0.0007/E)^2+(0.008/E^{1/4})^2$, where $E$ is in GeV.

The KLM is designed to detect $K_L$'s and muons. 
The KLM consists of 14 or 15 modules which contain of 47~mm thick iron plates and 44~mm thick slots instrumented with resistive plate counters (RPC).

In the physics analyses, we take advantage of the fact that the energy of each $B$ meson is precisely known from accelerator parameters. 
The following kinematic variables are commonly used to distinguish $B$ decay signals from backgrounds,
\begin{eqnarray}
\mb &\equiv &\sqrt{(\ebeam)^2 - (\sum {p}^*_i)^2},\nonumber\\
\dE &\equiv& \sum E^*_i - \ebeam,
\label{eq:it}
\end{eqnarray}
where $\ebeam$ is the beam energy in the CM frame,  ${p}^*_i$ and $E^*_i$ are the energies and momentum vectors of the $B$ candidate decay daughters in the CM frame.
The \Mb\ resolution is dominated by the resolution in
$\ebeam$ and is very narrow, typically 3 \Mevcsq.
This variable is useful to distinguish combinatorial backgrounds because of the good resolution.
The \DE\ variable peaks at zero for signal events and 
has a typical resolution of about 10 MeV.
This variable is useful to distinguish correlated backgrounds such as feed across and particle misidentification since missing or extra particles, or misidentified particles cause a shift in the \DE\ distribution.
The \Mb\ distribution still peaks in the signal region for these backgrounds.

In Monte Carlo simulation, the physics properties of an event are generated by the QQ event generator developed by the CLEO group.
The detector response is simulated using the CERN GEANT package.\cite{GEANT3}

A detailed description of the Belle detector can be found elsewhere.\cite{Belle}

\section{Measurement of Magnitude of CKM Matrix Elements}
The magnitude of the CKM matrix elements, $|\vcb|$ and $|\vub|$ can be measured by studying semileptonic $B$ decays.
Semileptonic decay is theoretically easier to understand since there
is no final state interactions between the lepton and hadron systems.
In the naive spectator model, the partial decay width for the inclusive semileptonic $B$ decay, $\bbar\to X\ell^-\nub$,\cite{charge-conjugate} can be expressed as:
\begin{equation}
\Gamma(\bbar\to X\ell^-\nub) = \frac{G_F^2m_b^5}{192\pi^3}(\gamma_{c}|\vcb|^2+\gamma_{u}|\vub|^2),
\end{equation}
where $\ell^-=e^-,\;\mu^-$ and $\gamma_{q}$ incorporates phase space and possible strong interaction effects.

Experimentally, the inclusive semileptonic branching fraction can be measured using the dilepton method introduced by ARGUS.\cite{dilepton}
In this method, high momentum lepton is required to tag the flavor of one $B$ meson and the other lepton (only electron is used) is used for the measurement.
Primary and secondary electrons are identified by the charge correlation between the electron and the tagged lepton.
The primary electron and the tagged lepton have opposite charge unless the two $B$ mesons have the same flavor due to $\bz$-$\bzb$ mixing.
The major backgrounds include secondary electrons from 
the decay chain $b\to c\to y\ell^-\nub$ ($y=s,d$) of the tagged $B$
meson and electrons from continuum $e^+e^-\to q \overline{q}$ events, which are suppressed by using angular correlations between the tag lepton and the electron in the CM frame.

\begin{figure}
\center{\resizebox{0.50\textwidth}{!}{\includegraphics{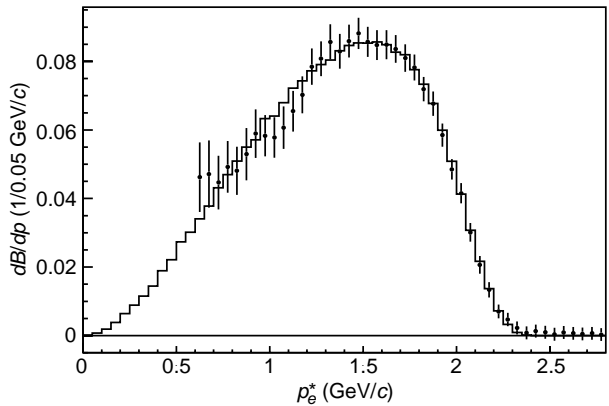}}}
\caption{Primary electron momentum spectrum. The histogram represents the fit result to the ISGW2 model.}
\label{fig:e-spectrum}
\end{figure}
\begin{table}
\caption{$\vcb$ determination through the inclusive semileptonic branching fraction.}\label{table:vcb-inclusive}
\begin{center}
\begin{tabular}{|c|c|c|} 
\hline 
Model  						& $|\vcb|$ ($\times10^{-2}$) \\ \hline
ACCMM\cite{ACCMM} 			& $4.1\pm0.1\pm0.4$\\ \hline
ISGW\cite{ISGW} 			& $4.0\pm0.1\pm4$\\ \hline
Shifman \etal\cite{shifman}	& $4.04\pm0.10\pm0.20$\\ \hline
Ball \etal\cite{ball} 		& $3.95\pm0.09\pm0.19$\\ \hline
\end{tabular}
\end{center}
\end{table}
Figure \ref{fig:e-spectrum} shows the spectrum for the primary electron, ${d\BF(b\to x\ell^- \nub )} /{dp}$, after subtracting the background and taking into account the effect of mixing.
The branching fraction is obtained by a fit to the momentum spectrum predicted by theoretical models.\cite{ACCMM,ISGW}
Model dependence is very small since the measurement covers a large portion of the spectrum.
Using a 5.1~\Fb\ data sample, the inclusive semileptonic branching fraction is measured to be
\begin{equation}
\BF(\bbar\to Xe^- \nub ) =  (10.86\pm0.14\pm0.47) \%,
\end{equation}
where the first error is statistical and the second error is systematic.
The systematic error is dominated by the uncertainty in the electron identification and kinematic selection efficiencies.
The $|\vcb|$ value is derived from the measured branching fraction and the model calculation of $\gamma_c$ ignoring the $|\vub|$ term.
Model dependence is relatively large as summarized in Table \ref{table:vcb-inclusive}.

\bigskip

The value of $|\vcb|$ can be determined with less model dependence using HQET (Heavy Quark Effective Theory).\cite{HQET}
In HQET, the partial $\bzdstlnu$ decay rate is expressed as\cite{HQET-FF}
\begin{eqnarray}
&&\frac{d\Gamma(\bzdstlnu)}{dy}=\frac{G_F^2}{48\pi^3}M_{\dstp}^2(M_{\bzb}-M_{\dstp})^2|\vcb|^2g(y)F(y)^2,
\end{eqnarray}
where $y\equiv { v}_{\bzb}\cdot{
v}_{\dstp}=(M_{\bzb}^2+M_{\dstp}^2-q^2)/(2M_{\bzb}M_{\dstp})$, ${
v}$ is the four-momentum divided by the particle mass 
and $q^2$ is the square of the four momentum transfer.
The form factor at $y=1$ (zero recoil) can be calculated with small theoretical error.
Experimentally, it can be extrapolated from the $y$ distribution using the following parameterization;\cite{HQET-FF}
\begin{eqnarray}
g(y)F(y)^2&=&\sqrt{y^2-1}(y+1)^2A_1(y)^2\tilde{R}(y),\nonumber \\
\tilde{R}(y)&=&2\frac{1-2yr+r^2}{(1-r)^2}(1+R_1(y)^2\frac{y-1}{y+1}) 
+\{1+(1-R_2(y))\frac{y-1}{1-r}\}^2,  
\end{eqnarray}
where $r=M_{\dstp}/M_{\bzb}$.
We have used a dispersion relation\cite{dispersion} to constrain the shape of the form factor,
\begin{eqnarray}
R_1(y)&\approx&R_1(1)-0.12x+0.05x^2, \nonumber\\
R_2(y)&\approx&R_2(1)-0.11x+0.06x^2, \\
A_1(y)&\approx&A_1(1)[1-8\rho^2z+(53\rho^2-15)z^2-(231\rho^2-91)z^3], \nonumber
\end{eqnarray}
where $x=y-1$ and $z=(\sqrt{y+1}-\sqrt{2})/(\sqrt{y+1}+\sqrt{2})$.

\begin{figure}
\center{\resizebox{0.60\textwidth}{!}{\includegraphics{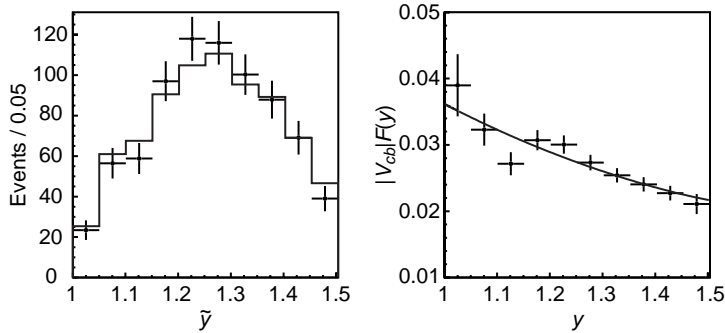}}}
\caption{The plot on the left shows the $\bzdstlnu$ raw yield as a function
of the measured $y$. The solid circles and the histogram represent
the data and the fit result, respectively. 
The plot on the right shows $|\vcb|F(y)$ where the data points are derived
from the yield by correcting for efficiency, smearing and all the factors in the differential decay rate.
The curve displays the fit result.}
\label{fig:y-Dstenu}
\end{figure}

Candidate $\bzdstlnu$ decays are selected by applying kinematic constraints on events with a electron and a $\dstarp\to \dz\pip$ $\dz\to\kpi$ decay chain.
The values of $|\vcb|F(1)$ and $\rho^2$ are extracted by a binned
minimum $\chi^2$ fit to the $y$ distribution after background
subtraction.
The results were obtained from a 10.8~\Fb\ data sample while fixing $R_1(1)=1.27$ and $R_2(1)=0.8$.\cite{HQET-FF}
Figure~\ref{fig:y-Dstenu} shows the $y$ distribution with the fit results.
The fit yields
\begin{eqnarray}
|\vcb|F(1)&=&(3.54\pm0.19\pm0.19)\times10^{-2},\nonumber\\
\rho^2&=&1.35\pm0.17\pm0.18.
\end{eqnarray}
Using $F(1)=0.913\pm0.042,$\cite{HQET-F1} $|\vcb|$ is determined to be
\begin{equation}
|\vcb| = (3.88\pm0.21\pm0.21\pm0.19)\times10^{-2}.
\end{equation}
where the third error is theoretical.
The systematic error is dominated by the uncertainty in the tracking efficiency for slow pions from $\dstp\to\dz\pip$ decay.

\bigskip

The magnitude of $\vub$ is one of the smallest and poorly measured
 parameters of the CKM matrix.
This is the key element needed to evaluate the consistency of the
 SM with the large $CP$ angle, 
$\sin2\phi_1=0.99\pm0.14\pm0.06$,\cite{sin2phi1} measured by Belle.
Exclusive semileptonic $b\to u\ell^-\nub$ decay is one of the most promising modes to determine $|\vub|$ provided that the form factors are reliably calculated by theoretical models.

The branching fractions for $\bzb\to\rho\ell^-\nub$ and $\bzb\to\pi\ell^-\nub$ have previously been measured by CLEO using ``neutrino reconstruction" technique.\cite{b2ulnu-CLEO}
In this method, the four momentum of the undetected neutrino is inferred from the missing momentum and energy in an event.
The missing momentum is a poor measure of the neutrino momentum, when the event has multiple neutrinos, missing or extra particles.
\begin{figure}
\center{\resizebox{0.60\textwidth}{!}{\includegraphics{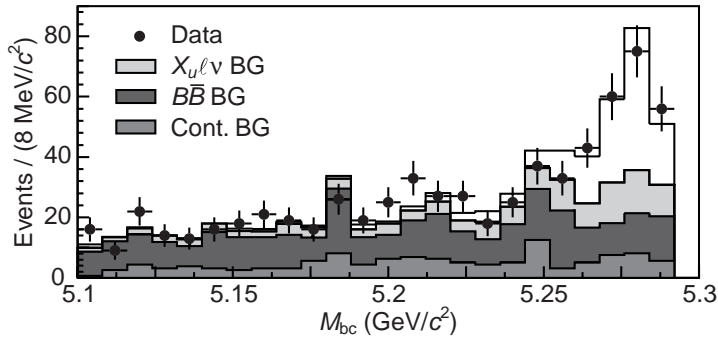}}}
\caption{\Mb\ distribution for $\bzb\to\pi^+\ell^-\nub$ decay. The unshaded histogram shows the signal component while the hatched histograms show various background components.}
\label{fig:pilnu}
\end{figure}
Such events are reduced by requirements on total charge and the number of leptons.
The missing mass-squared of the event in the CM frame, $(M_{\rm miss})^2=(E_{\rm miss}^*)^2-(p_{\rm miss}^*)^2$, is required to satisfy $|(M_{\rm miss})^2|<2$~(\Gevcsq)${}^2$ to further suppress events with poorly reconstructed neutrinos.
The major background from $b\to c\ell^-\nub$ is reduced by requiring $|p^*_\ell|>1.2$~\Gevc, $|p^*_\ell|+|p^*_\pi|>3.1$~\Gevc\ and other kinematic constraints.
We require $|\dE|<0.3$~GeV to reduce feed-across as well as combinatorial backgrounds.
The signal yield is extracted by a fit to the \Mb\ distribution.
A Monte Carlo (MC) simulation is used to provide both signal and background shapes.
We find $107\pm16$ $\bzb\to\pi^+\ell^-\nub$ events in a 21.3~\Fb\ data sample as shown in Figure~\ref{fig:pilnu}, corresponding to a branching fraction of $\BF(\bzb\to\pi^+\ell^-\nub)=(1.24\pm0.20\pm0.26)\times10^{-4}$.
The systematic error due to uncertainty of the neutrino finding is dominant.
Further studies are required to extract $|\vub|$ from this measurement.

\section{Toward $\phi_i$ Measurements}
The main motivation of the Belle experiment is to measure all the $CP$ angles ($\phi_i$).
Such measurements are essential to probe new physics and
overconstrain the CKM matrix.
We report some measurements and observations of new decay modes 
which demonstrate our capabilities to measure these $CP$ angles.

Doubly-charmed decay modes $\bzb\to \dstp\dminus,\;\dplus\dstm$ (referred as $D^{*\pm}D^{\mp}$ hereafter), $\dstp\dstm$ can be used to measure\cite{DD-theory} $\phi_1$ in addition to the gold plated mode $\bz\to\psi\kz$.
The $CP$ asymmetry parameter measured in this mode could deviate from the expected value due to a sizable penguin contribution, which may provide evidence for new physics.
The $\bzb\to D^{*\pm}D^{\mp}$ decay mode has not been observed so far.

Candidates are selected using the decay chains, $\dminus\to\kp\pim\pim$ and $\dstp\to\dz\pip$, $\dz\to\kpi$, $\kpipiz$, $\kpipipi$.
We also use the $\dz\to\ks\pip\pim,\;\ks\to\pip\pim$ decay chain for the $\bzb\to D^{*+}D^{*-}$ mode.
The signal yield is extracted by a maximum likelihood (ML) fit to the \Mb\ distribution for
events within the \DE\ signal region ($|\dE|<20$ MeV).
A Gaussian and an ARGUS functions\cite{ARGUS} are used to represent
the signal and background shapes.
\begin{figure}
\vspace*{0.2cm}
{\footnotesize \hspace*{4.4cm}$\bzb\to D^{*\pm}D^{\mp}$ \hspace*{2.5cm}$\bzb\to D^{*+}D^{*-}$}
\vspace*{-1.1cm}
\center{\resizebox{0.60\textwidth}{!}{\includegraphics{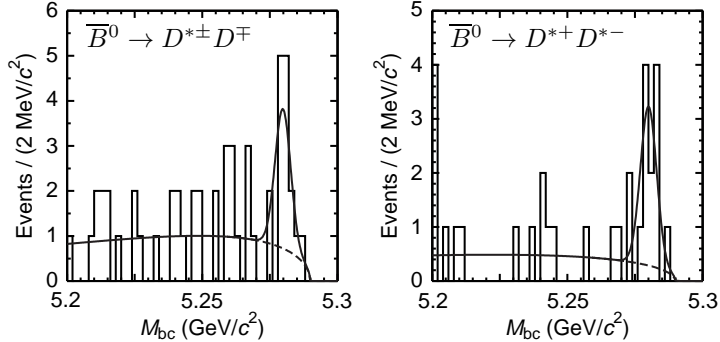}}}
\caption{\Mb\ distribution for $\bzb\to D^{*\pm}D^{\mp}$ and $\bzb\to D^{*+}D^{*-}$ modes. The solid line shows the sum of signal and background while the dashed line shows background.}
\label{fig:DD}
\end{figure}
The fits yield $11.2\pm4.0$ $\bzb\to D^{*\pm}D^{\mp}$ events and $11.0\pm3.7$ $\bzb\to\dstp\dstm$ events in a 21.3~\Fb\ data sample with statistical significances of $4.1\sigma$ and $5.0\sigma$, respectively. The statistical significance is defined as $\sqrt{-2\ln({\cal L}_0/{\cal L}_{\rm max})}$, where ${\cal L}_{\rm max}$ is the maximum likelihood and ${\cal L}_0$ is the likelihood values at zero signal yield.
Figure \ref{fig:DD} shows the \Mb\ distributions with the fit results.
The branching fractions are measured to be 
\begin{eqnarray}
\BF(\bzb\to D^{*\pm}D^{\mp})&=&(1.04\pm0.38\pm0.22)\times10^{-3},\nonumber\\
\BF(\bzb\to D^{*+}D^{*-})&=&(1.21\pm0.41\pm0.27)\times10^{-3}.
\end{eqnarray}
The systematic error is dominated by the uncertainties in tracking efficiency and fit method dependence.

\bigskip

The $\bzb\to D^{*\pm}D^{\mp}$ decay mode is also observed using a $\dstp$ partial reconstruction technique.
In this method, after the $\dstp\to\dz\pip$ decay 
we do not reconstruct the subsequent $\dz$ decay in order
to increase the overall detection efficiency.
Charmed mesons from $\bzb\to D^{*\pm}D^{\mp}$ are almost back-to-back in the CM frame since $B$ mesons are produced almost at rest.
A slow pion from the $\dstp\to\dz\pip$ decay approximately retains the momentum direction of the parent $\dstp$ because of the small energy release in this decay.
Thus the angle $\alpha$ between the slow pion and the $\dminus$ are almost back-to-back and can be employed as a signature.
However, the partial reconstruction method suffers from a relatively large background.
\begin{figure}
\vspace*{0.1cm}
{\footnotesize \hspace*{4.7cm}No lepton tag \hspace*{2.6cm}Lepton tag}
\vspace*{-1.0cm}
\center{\resizebox{0.60\textwidth}{!}{\includegraphics{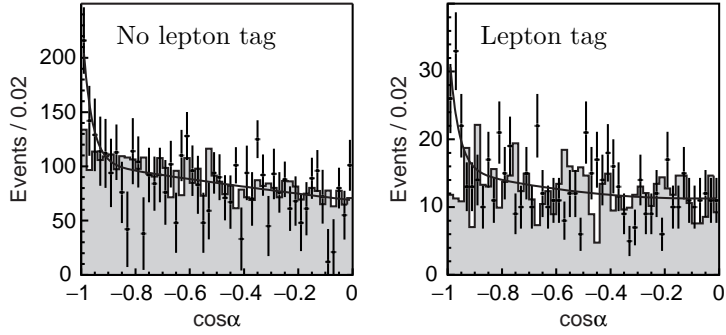}}}
\caption{$\cos\alpha$ distribution for partially reconstructed $\bzb\to D^{*\pm}D^{\mp}$ decay.}
\label{fig:partial-DDst}
\end{figure}
In particular, continuum $e^+e^-\to\ccbar$ events can produce
$\dstp$ and $\dminus$ in a back-to-back configuration, which results in similar angular correlation.
The partial reconstruction method also introduces a difficulty for the $\phi_1$ measurement since the particles which are not used for the reconstruction complicate the vertex reconstruction and the flavor tagging of the accompanying $B$ meson.
These problems can be alleviated by requiring a high momentum lepton ($p^*_\ell>1.1$ \Gevc) in the event.
This requirement heavily suppress the continuum background, and at the same time, provides a clean flavor tag and a good vertex reconstruction for the $\phi_1$ measurement.
The remaining backgrounds can be further suppressed by exploiting the angular correlation between the tag lepton and $\dminus$.
Figure~\ref{fig:partial-DDst} shows the $\cos\alpha$ distribution observed in a 21.3~\Fb\ data sample.

A fit to the $\cos\alpha$ distribution for the lepton tagged sample
yields $35.8\pm11.3$ signal events.
The signal and background shapes are obtained from the MC simulation except for the background from Cabbibo-favored $\bzb\to\dstp D_s^-$ decays for which the data is used to estimate the amount and the shape of the background.
The branching fractions are measured to be 
\begin{eqnarray}
\BF(\bzb&\to& D^{*\pm}D^{\mp})=(1.78\pm0.56^{+0.75}_{-0.63})\times10^{-3}.
\end{eqnarray}
Uncertainties in the background shape are the
 dominant source of the systematic error.
This measurement clearly demonstrates that the partial
 reconstruction technique can be used for a $\phi_1$ measurement
 with $\bzb\to D^{*\pm}D^{\mp}$ and $\bzb\to D^{*+}D^{*-}$ decays.

\bigskip

The gluonic penguin decay mode $\bzb\to\phi\ks$ can also provide an independent measurement of the $CP$ angle, $\phi_1$.\cite{phiK-theory}
This decay mode proceeds through a $b\to ss\overline{s}$ transition, which is forbidden at the tree level in the SM, but are induced by second order loop diagrams (penguin or box diagrams).
The $CP$ asymmetry parameter measured in this mode is of special
interest since it is sensitive to the possible exchange of non-SM particles in the loop\cite{kphi-non-SM} and may deviate from the expected value. 

The $\bzb\to\phi\ks$ candidates
are reconstructed via $\phi\to\kp\km$ and $\ks\to\pip\pim$ decays.
The dominant background arises from continuum events.
Event shape variables are used to suppress the continuum background.
The most powerful variable is the cosine of the angle between the $B$ candidate thrust axis and the thrust axis of the rest the event ($\cos\thrust$).
The $\cos\thrust$ distribution for the signal is flat while it is peaked at $\pm1$ for continuum events.
This variable is combined with other variables such as the $B$ flight direction and the helicity angle for pseudoscalar-vector final states ($\phi\ks,\;\phi\km$) using a likelihood ratio ${\cal LR}={\cal L}_S/({\cal L}_S+{\cal L}_B)$ where ${\cal L}_S$ and ${\cal L}_B$ denote the signal and background likelihoods.
The likelihood is the product of probability density function in each of the discriminating variables.

\begin{figure}
\vspace*{0.1cm}
\hspace*{4.5cm}{\footnotesize $\bzb\to\phi\ks$}
\vspace*{-1.0cm}
\center{\resizebox{0.60\textwidth}{!}{\includegraphics{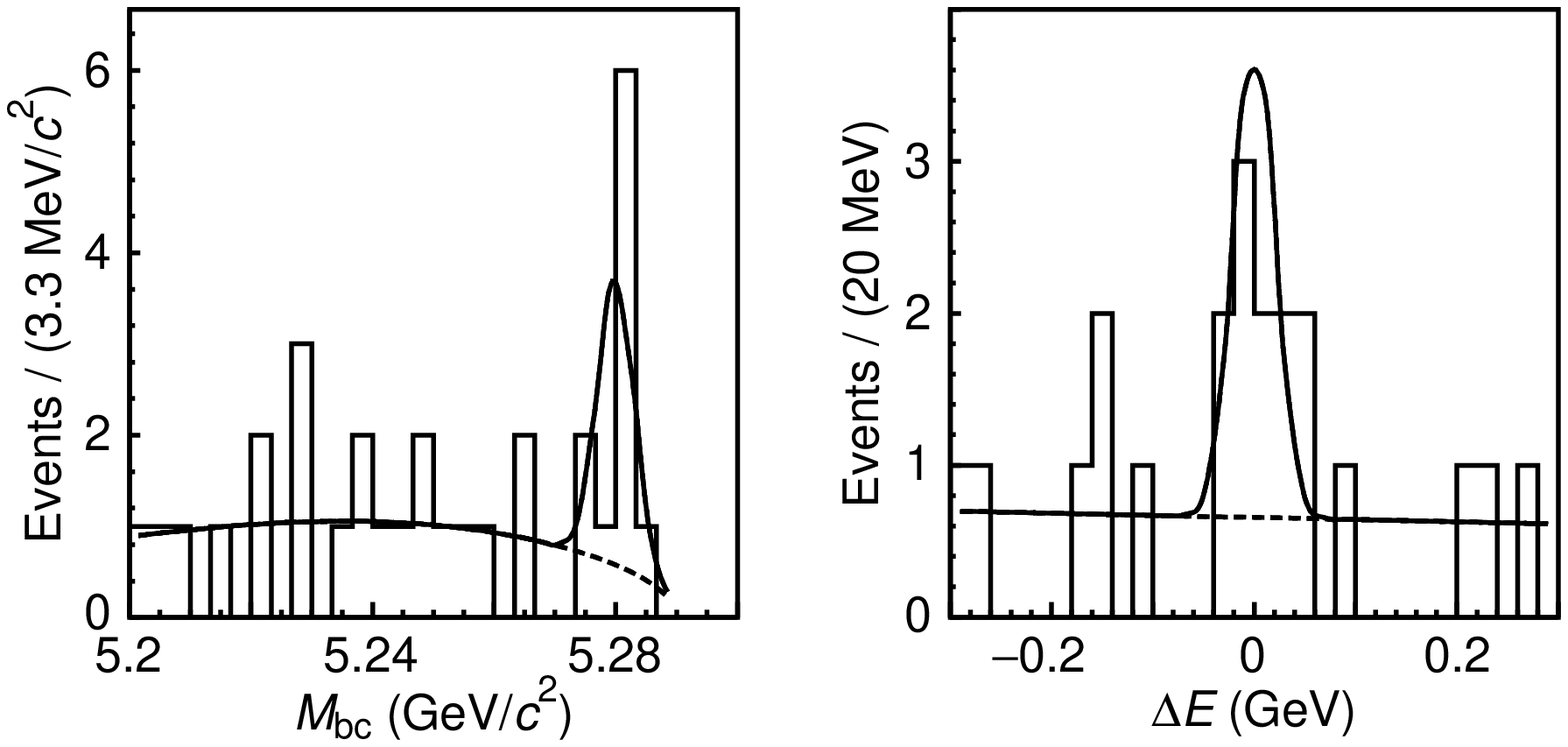}}}
\caption{\Mb\ and \DE\ distributions for $\bzb\to\phi\ks$ decay.}
\label{fig:phiK}
\end{figure}

An extended unbinned ML fit is performed in \DE\ and \Mb\ simultaneously to extract the signal yield.
In the extended ML fit, the sum of the signal and background yield is allowed to be different from the total number of the event in the fit.
The signal shape is represented by Gaussian function in both \DE\ and \Mb.
The background shape is represented by a linear function in \DE\ and an ARGUS function in \Mb.
The fit yields $8.0^{+3.5}_{-2.8}$ $\bzb\to\phi\ks$ events with a statistical significance of $4.2\sigma$ in a 21.3~\Fb\ data sample.
Figure \ref{fig:phiK} shows the \Mb\ and \DE\ distributions with the projection of the fit result.
We have also observed other $\bbar\to\phi\kbar$ modes and obtained the branching fractions as
\begin{eqnarray}
\BF(\bzb\to\phi\kzb)&=&(0.89^{+0.34}_{-0.27}\pm0.10)\times10^{-5},\nonumber\\
\BF(\bzb\to\phi\kstzb)&=&(1.30^{+0.64}_{-0.52}\pm0.21)\times10^{-5},\\
\BF(\bm\to\phi\km)&=&(1.12^{+0.22}_{-0.20}\pm0.14)\times10^{-5}.\nonumber
\end{eqnarray}
Here, $\kstzb$ refers to the $\kstzb(892)$. The dominant systematic errors come from uncertainties due to tracking and $\ks$ reconstruction.

\bigskip

Charmless $B$ decay $\bzb\to\pim\pip$ is one of the most promising modes to measure the $CP$ angle $\phi_2$.
We also need to study $\bz\to\piz\piz$ and $\bm\to\pim\piz$ decays
to disentangle the effect of penguins.\cite{penguin-pollution}
Experimentally, $K/\pi$ separation at high momentum is the key to distinguish the $\pi\pi$ mode from the $K\pi$ mode.
The ACC detector in the Belle plays an essential role in this regard.
The $K/\pi$ separation capability is measured using kinematicaly selected $\dstp\to\dz\pip,\;\dz\to\km\pip$ decays in the data.
The efficiency is measured to be 92\% for pion and 85\% for kaon.
The misidentification probability is measured to be 4\% for pion (true pion fakes kaon) and 10\% for kaon.
The dominant background comes from continuum events.
The Super-Fox-Wolfram (SFW) variable\cite{SFW} which is an extension of the normalized Fox-Wolfram moments\cite{FW} is employed to suppress the continuum background in addition to the $\cos\thrust$.
These event shape variables are combined into a Fisher discriminant\cite{Fisher} since they are correlated.
The resulting Fisher discriminant is combined with the $B$ flight direction using a likelihood ratio.
This selection rejects 95\% of the continuum background while retaining 40\% to 50\% of the signal.

\begin{figure}
\vspace*{0.1cm}
{\footnotesize \hspace*{5.7cm}$\bzb\to\pim\pip$}
\vspace*{-0.9cm}
\center{\resizebox{0.50\textwidth}{!}{\includegraphics{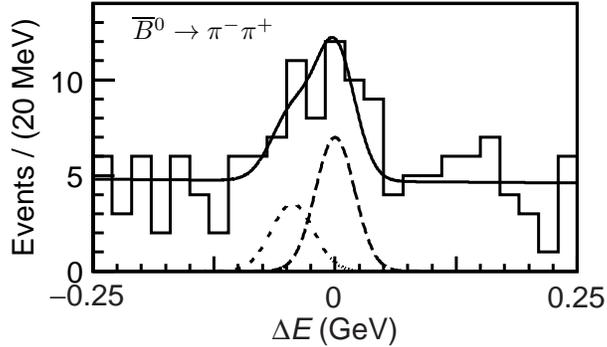}}}
\caption{\DE\ distribution for $\bzb\to\pim\pip$ decay. The fit function and its signal and cross-talk components are shown by the solid, dashed and dotted curve, respectively.}
\label{fig:dE-pipi}
\end{figure}

The signal yield for $\bzb\to\pim\pip$ mode is determined from a fit to the \DE\ distribution where there is kinematic separation between $\pim\pip$ and $\km\pip$ modes.
The signal shape is modeled by a Gaussian function while the background shape is modeled by a linear function and a Gaussian function.
The Gaussian function is introduced to account for the background from $K\pi$ mode.
The mean value of this background is shifted by about $-50$~MeV
since the kaon is misidentified as pion.
The normalization of all components are free parameters in the fits.
The Gaussian background yield extracted from the fit is consistent with the misidentification probability mentioned above.
Figure \ref{fig:dE-pipi} shows the \DE\ distribution for $\bzb\to\pim\pip$ mode along with the fit result.
The signal yield for $\bm\to\pim\piz$ mode is determined from the simultaneous fit to the \DE\ and \Mb\ distribution because of the long tail for the \DE\ signal shape.
We find $17.7^{+7.1}_{-6.4}$ events for the $\bzb\to\pim\pip$ mode
and $10.4^{+5.1}_{-4.3}$ events for the 
$\bm\to\pim\piz$ mode in 10.4~\Fb\ of data.
The branching fractions are measured to be 
\begin{eqnarray}
\BF(\bzb\to\pim\pip)&=&(0.56^{+0.23}_{-0.20}{}^{+0.04}_{-0.05})\times10^{-5},\nonumber\\
\BF(\bm\to\pim\piz)&=&(0.78^{+0.38}_{-0.32}{}^{+0.08}_{-0.12})\times10^{-5}.
\end{eqnarray}

\bigskip

The interference between $\bm\to\dz\km$ ($b\to c\overline{u} s$) and
$\bm\to\dzb\km$ ($b\to u\overline{c} s$) decays provides a
theoretically clean determination of the angle $\phi_3$.\cite{B2DCPK,B2DK}
When $\dz$ and $\dzb$ decays into a common $CP$ eigenstate mode, the $CP$ angle $\phi_3$ can be related to the following observables\cite{B2DCPK} assuming no $\dz$-$\dzb$ mixing;
\begin{eqnarray}
A&\equiv&\frac{\Gamma(\bm\to\dcp\km)-\Gamma(\bplus\to\dcp\kp)}{\Gamma(\bm\to\dcp\km)+\Gamma(\bplus\to\dcp\kp)}\nonumber\\
&\approx&2\xi_f r\sin\delta\sin\phi_3,\nonumber\\
R&\equiv&\rho'/\rho=1+r^2+2\xi_f r\cos\delta\cos\phi_3,\\
\rho'&\equiv&\frac{\Gamma(\bm\to\dcp\km)+\Gamma(\bplus\to\dcp\kp)}{\Gamma(\bm\to\dcp\pim)+\Gamma(\bplus\to\dcp\pip)},\nonumber\\
\rho&\equiv&\frac{\Gamma(\bm\to\dz\km)+\Gamma(\bplus\to\dzb\kp)}{\Gamma(\bm\to\dz\pim)+\Gamma(\bplus\to\dzb\pip)},\nonumber\\
r&\equiv&A(\bm\to\dzb\km)/A(\bm\to\dz\km),\nonumber
\end{eqnarray}
where $\dcp$ is the $CP$ eigenstate of the $\dz$ meson, $\delta$ is the strong phase-difference between $\bm\to\dzb\km$ and $\bm\to\dz\km$, and $\xi_f$ is the $CP$ eigenvalue of $\dcp$.

\begin{figure}
{\footnotesize \hspace*{5.5cm}$\bm\to \dz\km$\vspace*{-0.1cm}\\
\hspace*{8.7cm}$\dz\to\kpi$\vspace*{1.20cm} \\
\hspace*{8.7cm}$\dz\to\km\kp$\vspace*{1.20cm} \\
\hspace*{8.7cm}$\dz\to\pim\pip$}
\vspace*{-4.5cm}
\center{\resizebox{0.40\textwidth}{!}{\includegraphics{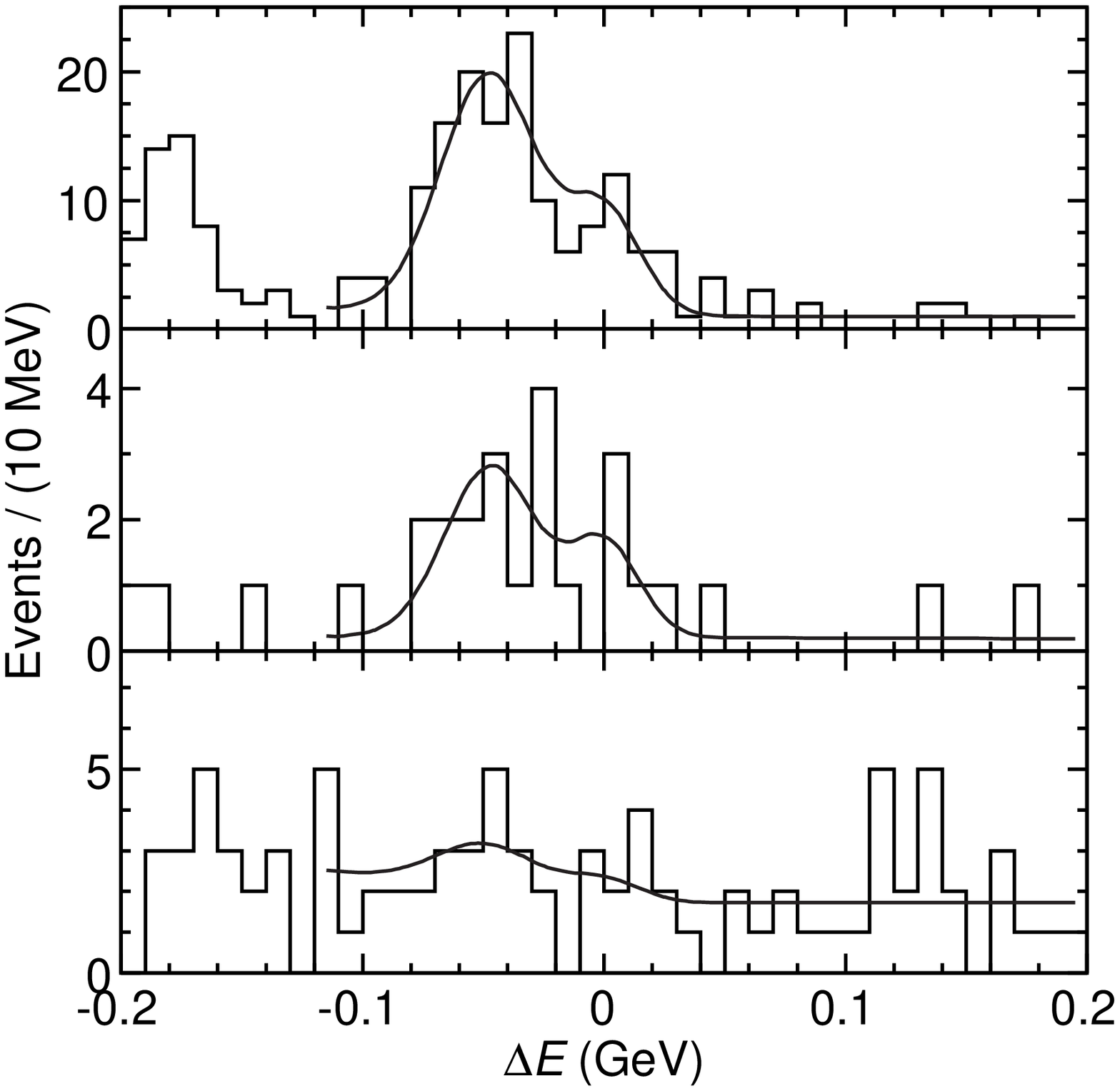}}}
\caption{\DE\ distribution for for $\bm\to \dz\km$ decay with subsequent $\dz$ decay into $\kpi$, $\km\kp$ and $\pim\pip$. The fit function is described in the text.}
\label{fig:dE-DK}
\end{figure}

We have studied $\bm\to \dz\km$ decays where $\dz$ decays into $\kpi$ or into a $CP=+1$ eigenstate ($\km\kp$, $\pim\pip$).
Continuum events are the dominant background and are reduced by a selection based on event shape variables.
Tight particle identification is applied to reduce the background from $\bm\to\dz\pim$ decays.
The signal yields are extracted from a fit to the \DE\ distribution 
that accounts for the remaining $\bm\to\dz\pim$ background.
The background function includes an ARGUS function to model the combinatorial background and a Gaussian function to model the $\bm\to\dz\pim$ background.
Figure \ref{fig:dE-DK} shows the \DE\ distributions with the fit results.
It should be noted that peak position for the signal is shifted by
$-49$~MeV since \DE\ is calculated with the assumption of a 
pion mass for the prompt kaon.
We find $1278\pm37$, $114\pm11$ and $70\pm11$ $\bm\to\dz\pim$ events and $85\pm10$, $12.3\pm3.9$ and $4.9\pm5.4$ $\bm\to \dz\km$ events with subsequent $\dz$ decays into $\kpi$, $\km\kp$ and $\pim\pip$ modes, respectively, in 21.3~\Fb\ of data.
The statistical significance of the $\bm\to \dz\km,\;\dz\to\km\kp$ signal
is $4.3\sigma$.
These results gives
\begin{eqnarray}
A&=&0.04^{+0.40}_{-0.35}\pm0.15,\nonumber\\
R&=&1.39\pm0.53\pm0.26.
\end{eqnarray}
This measurement is an important first step toward a $\phi_3$ measurement.

\section{Charmless Hadronic Decays}

Charmless hadronic $B$ decays proceed primarily through $b\to u$
tree diagrams and $b\to s$ penguin diagrams, which provide a 
rich ground to search for direct $CP$ violation,
although it is theoretically difficult to relate 
the $CP$ angles to the $CP$ symmetries measured in these decay modes.
In addition, there are possibilities to extract $CP$ angles from ratios of $B\to K\pi$ decay rates.\cite{Kpi-theory}
Furthermore, the involvement of the penguin diagrams makes 
these decays sensitive to the effect of new particles at higher mass
scales in the loop.

\bigskip
It is a theoretical challenge to explain the unexpectedly large branching fractions for $B\to\eta' K$ and $B\to\eta K^*$ decays reported by CLEO.\cite{etaK-CLEO}
Since it
 may suggest new physics beyond the SM, it is necessary to experimentally confirm the CLEO results and search for direct $CP$ violation which may be enhanced by new physics.
The large branching fraction for $\bzb\to\eta'\ks$ also opens the
 possibility of measuring mixing-induced $CP$ violation.

\begin{figure}
{\footnotesize
\vspace*{0.15cm}\hspace*{4.5cm}$\bzb\to\eta'\ks$}
\vspace*{-1.05cm}
\center{\resizebox{0.60\textwidth}{!}{\includegraphics{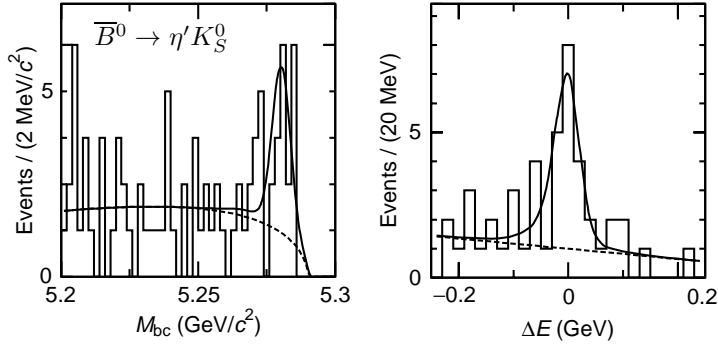}}}
\caption{\Mb\ and \DE\ distributions for $\bzb\to\eta'\ks$ mode. Solid and dashed curves show the projections of the signal$+$background and background functions.}
\label{fig:etapK}
\end{figure}

We analyze $\bm\to\eta'\km$, $\bzb\to\eta'\ks$ and
$\bzb\to\eta\kstzb$ decay through the decay chains $\eta'\to\eta\pip\pim\;(\eta\to\gamma\gamma)$, $\eta'\to\rho^0\gamma\;(\rho^0\to\pip\pim$), $\ks\to\pip\pim$ and $\kstzb\to\km\pip$.
We also use the $\eta\to\pip\pim\piz$ decay for the $\bzb\to\eta\kstzb$ analysis.
After background suppression using variables such as $\cos\thrust$, SFW variable, $B$ flight direction and helicity angle for $\bzb\to\eta\kstzb$ mode, we perform extended ML fits to both \Mb\ and \DE\ to extract signal yields.
We observe 71.4 $\bm\to\eta'\km$ and 16.5 $\bzb\to\eta'\ks$ events with statistical significances of $12.0\sigma$ and $5.4\sigma$, respectively, in 10.1~\Fb\ of data, and also 22.1 $\bzb\to\eta\kstzb$ events at $5.1\sigma$ in 21.3~\Fb\ of data.
Figure~\ref{fig:etapK} shows the \Mb\ and \DE\ distributions for $\bzb\to\eta'\ks$ mode with the projections of the fit function.
The branching fractions are measured to be
\begin{eqnarray}
\BF(\bm\to\eta'\km)&=&(79^{+12}_{-11}\pm9)\times 10^{-6},\nonumber\\
\BF(\bzb\to\eta'\kzb)&=&(55^{+19}_{-16}\pm8)\times 10^{-6},\nonumber\\
\BF(\bzb\to\eta\kstzb)&=&(21.2^{+5.4}_{-4.7}\pm2.0)\times 10^{-6}.
\end{eqnarray}
These results confirms the large branching fractions in these modes observed by CLEO.
In addition, we have measured the decay asymmetry for the
$\bm\to\eta'\km$ mode,
\begin{eqnarray}
A_{CP}&\equiv&\frac{\Gamma(\bm\to\eta'\km)-\Gamma(\bplus\to\eta'\kp)}{\Gamma(\bm\to\eta'\km)+\Gamma(\bplus\to\eta'\kp)}\nonumber\\
&=&+0.06\pm0.15\pm0.01.
\end{eqnarray}
The result is consistent with no direct $CP$ violation and with the SM prediction.

\bigskip

Three-body charmless decays $\bm\to\km h^+h^-$ ($h$ refers to a
charged kaon or pion) proceed through variety of diagrams such as
$b\to u\overline{u}d$, $b\to c\overline{c}s$ and $b\to
s\overline{s}s$, which provide an excellent environment 
to search for $CP$ violation due the interference of these amplitudes.
In particular, a great deal of attention is paid to the interference
between the former two amplitudes because it is sensitive to
 the $CP$ angle $\phi_3$.
The $\bm\to\pim\pip\pim$ decay is a good example since the 
non-resonant mode can interfere with the decay $\bm\to\chi_{c0}\pim,\;\chi_{c0}\to\pim\pip$.\cite{chicpi}

In this analysis, we reconstruct the decays $\bm\to\km h^+h^-$ without any assumption on the intermediate hadronic resonance.
As in other rare $B$ decay modes, continuum events are the dominant background source and are suppressed using various event shape and kinematic variables.
In the $\bm\to\km\pip\pim$ mode, we have large backgrounds from $\bm\to\dz\pim,\;\dz\to\km\pip$ and $\bm\to \km\psi^{(')},\;\psi^{(')}\to\mu^+\mu^-$ ($\psi^{(')}$ refers to $J/\psi$ and $\psi(2\mathrm{S})$) where muons are misidentified as pions.
These backgrounds are suppressed by requiring $|M_{\kpi}-M_{\dz}|>0.1$~\Gevcsq, $|M_{h^+h^-}-M_{J/\psi}|>0.07$~\Gevcsq\ and $|M_{h^+h^-}-M_{\psi(2\mathrm{S})}|>0.05$~\Gevcsq.
In the $\bm\to\km\kp\km$ mode, the background from the $\bm\to\dz\km,\;\dz\to\km\kp$ decay is rejected by requiring $|M_{\km\kp}-M_{\dz}|>0.025$~\Gevcsq.
Signal yields are obtained from fits to the \DE\ distributions.
\begin{figure}
{\footnotesize \vspace*{0.15cm}\hspace*{4.2cm}$\bm$$\to$$\km\pip\pim$\hspace*{2.6cm}$\bm$$\to$$\km\kp\km$}
\vspace*{-1.05cm}
\center{\resizebox{0.60\textwidth}{!}{\includegraphics{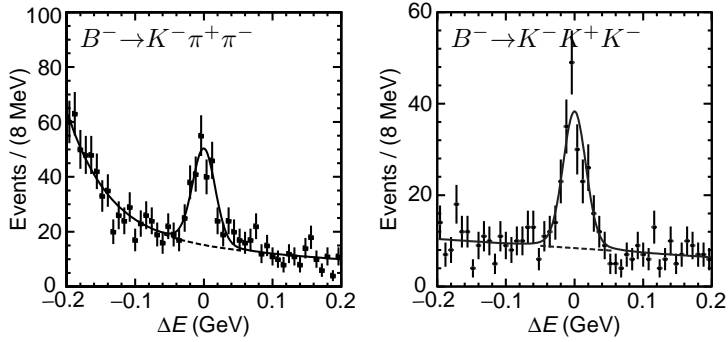}}}
\caption{\DE\ distributions for $\bm\to\km\pip\pim$ and $\bm\to\km\kp\km$ modes. 
Points are data, histograms are backgrounds estimated by the MC, and solid and dashed curves show the signal$+$background and background functions.}
\label{fig:Khh}
\end{figure}
We find $177\pm20$ $\bm\to\km\pip\pim$ events and $162\pm16$ $\bm\to\km\kp\km$ events in a 21.3 \Fb\ data sample, which corresponds to branching fractions;
\begin{eqnarray}
\BF(\bm\to\km\pip\pim)&=&(58.5\pm7.1\pm8.8)\times 10^{-6}, \nonumber\\
\BF(\bm\to\km\kp\km)&=&(37.0\pm3.9\pm4.4)\times 10^{-6}.
\end{eqnarray}
Figure \ref{fig:Khh} shows the \DE\ distribution for the $\bm\to\km\pip\pim$ and the $\bm\to\km\kp\km$ decays after the background suppression.
Our result is higher than the upper limit, $\BF(\bm\to\km\pip\pim)<28\times 10^{-6}$, reported by CLEO.\cite{Kpipi-CLEO}
The CLEO's analysis required the mass for any pair of particles to
be above 2~\Gevcsq, which effectively eliminates the low mass
resonances that are found to dominate the signal.

Further studies of intermediate resonant states of these decays are
made with a Dalitz plot style analysis.
Clear contributions from $\bm\to\kstzb\pim$, $\kstzb\to\km\pip$ and $\bm\to\km\fz(980)$, $\fz\to\pip\pim$, and no $\km\rho^0$ signal are observed in the $\bm\to\km\pip\pim$ decay.
We also find broad resonances in $\km\pip$ and $\pip\pim$ mass around 1.4~\Gevcsq\ and 1.3~\Gevcsq, respectively
In the $\bm\to\km\kp\km$ mode, we find a clear $\phi(1020)$ resonance and a broad $\kp\km$ enhancement around 1.6~\Gevcsq.
Larger statistics is required to identify these resonances using a Dalitz plot analysis with interference terms between the resonances.

\begin{figure}
{\footnotesize \vspace*{0.15cm}\hspace*{4.35cm}$\bm$$\to$$\km\pip\pim$\hspace*{2.6cm}$\bm$$\to$$\km\kp\km$}
\vspace*{-1.15cm}
\center{\resizebox{0.60\textwidth}{!}{\includegraphics{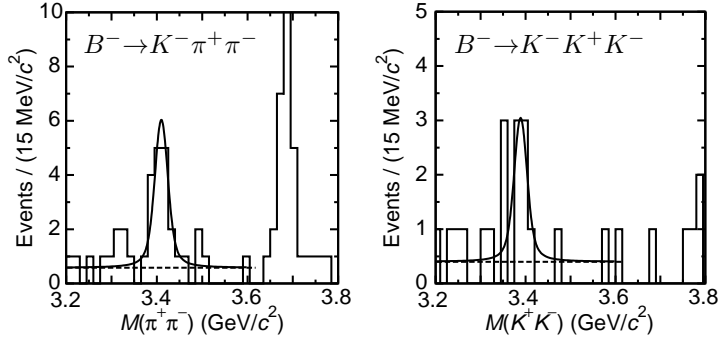}}}
\caption{The $\pip\pim$ mass distribution in the $\bm\to\km\pip\pim$ decay and the $\kp\km$ mass distribution in the $\km\kp\km$ decay. The signal function is modeled by a Breit-Wigner function convolved with a Gaussian.}
\label{fig:chic0}
\end{figure}

We also find clear signals for $\bm\to\chi_{c0}\km$, $\chi_{c0}\to\pip\pim$, $\kp\km$ in $\bm\to\km\pip\pim$, $\km\kp\km$ decays.
Figure \ref{fig:chic0} shows the $\pip\pim$ and $\kp\km$ mass distributions in $\bm\to\km\pip\pim$, $\km\kp\km$ decays.
The fits to these distributions yield $15.5^{+5.3}_{-4.6}$ events in the $\pip\pim$ mode and $7.7^{+3.9}_{-3.1}$ events in the $\kp\km$ mode at statistical significances of $4.8\sigma$ and $3.2\sigma$, respectively.
The fit also gives a sizable peak shift in the $\kp\km$ mass
spectrum, which may be
due to interferences from other $\km\kp\km$ final states.
Because of this uncertainty, we determine a branching fraction from the $\pip\pim$ mode only,
\begin{eqnarray}
\BF(\bm\to\chi_{c0}\km)&=&(8.0^{+2.8}_{-2.4}\pm1.9\pm1.1)\times 10^{-6},
\end{eqnarray}
where the third error is due to the uncertainty of the $\chi_{c0}\to\pip\pim$ branching fraction.
This observation provides evidence for a significant nonfactorizable
contribution in $B$ to charmonium decay process and suggests the
existence of a sizable $\bm\to\chi_{c0}\pim$ decay, 
which could be used for a $\phi_3$ measurement.

\section{Color-Suppressed Decays}
Color-suppressed $B$ decays such as $\bzb\to D^{(*)0} h^0$, where $h^0$ refers a light neutral mesons, proceed through an internal spectator diagram which is suppressed by color-matching.
Since the branching fractions for these modes are expected to be very small\cite{D0h0-theory} ($<10^{-4}$),
studies of color-suppressed decay could provide useful information on final state interactions and hadronic $B$-decay models.

In this study, $\bzb\to D^{(*)0} h^0$ decays are reconstructed through $\dstz\to\dz h^0$, $\dz\to\kpi$, $\km\pip\piz$, $\km\pip\pim\pip$ where $h^0$ is reconstructed through $\piz\to\gamma\gamma$, $\eta\to\gamma\gamma$, $\pip\pim\piz$ and $\omega\to\pip\pim\piz$ modes.
In addition to the continuum background which is suppressed by event shape variables and helicity angles, we pay attention to the background from color-favored decays.
The $\bzb\to\dstp\rho^-$ mode can give the same final state as $\dz\eta$ and $\dz\omega$ modes.
Such background is mostly removed by the $\eta$ and $\omega$ mass
constraints and its contribution can be evaluated from the $\eta$
and $\omega$ mass sidebands.
The $\bm\to D^{(*)0} \rho^-$ decays can contaminate the $\bzb\to D^{(*)0}\piz$ mode if the momentum of the $\pim$ from $\rho^-$ is very small.
These backgrounds are reduced by rejecting the events that can be reconstructed as $\bm\to D^{(*)0} \rho^-$.
These events also have \DE\ values lower than the signal due to the missing pion.

\begin{figure}
{\footnotesize \vspace*{0.25cm}\hspace*{4.3cm}$\bzb\to\dz\piz$\hspace*{3.2cm}$\bzb\to\dz\omega$}
\vspace*{-1.2cm}
\center{\resizebox{0.60\textwidth}{!}{\includegraphics{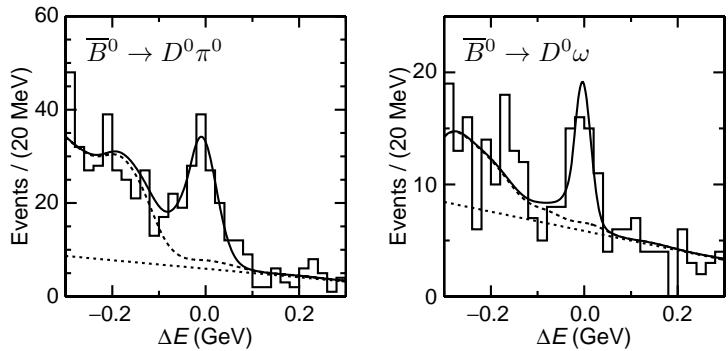}}}
\caption{\DE\ distributions for $\bzb\to\dz\piz$ and $\bzb\to\dz\omega$ modes. 
The solid lines show the fit functions, and the dashed lines show the sum of the feed-across and the combinatorial background, with the latter shown separately as the dotted lines.}
\label{fig:D0h0}
\end{figure}

Signal yields are extracted by fits to the \DE\ distributions taking 
into account the backgrounds from color-favored modes.
We observe $126\pm16$ events in $\dz\piz$ mode, $26.4^{+7.7}_{-7.1}$
events in the $\dstz\piz$ mode, $22.1^{+7.0}_{-6.3}$ events in the
$\dz\eta$ mode and $32.5^{+9.4}_{-8.6}$ events in the $\dz\omega$
mode in a 21.3~\Fb\ data sample corresponding to statistical
significances  of $9.3\sigma$, $4.1\sigma$, $4.2\sigma$ and
$4.4\sigma$, respectively.
Figure~\ref{fig:D0h0} shows the \DE\ distributions for the 
$\dz\piz$ and $\dz\omega$ modes along with the fit results.
The branching fractions are measured to be
\begin{eqnarray}
\BF(\bzb\to\dz\piz)&=&(3.1\pm0.4\pm0.5)\times 10^{-4},\nonumber\\
\BF(\bzb\to\dstz\piz)&=&(2.7^{+0.8}_{-0.7}{}^{+0.5}_{-0.6})\times 10^{-4},\nonumber\\
\BF(\bzb\to\dz\eta)&=&(1.4^{+0.5}_{-0.4}\pm0.3)\times 10^{-4},\\
\BF(\bzb\to\dz\omega)&=&(1.8\pm0.5^{+0.4}_{-0.3})\times 10^{-4}.\nonumber
\end{eqnarray}
These results are consistently higher than the predictions, $(0.5\sim1.0)\times10^{-4}$, by a factorization model,\cite{D0h0-theory} which could indicate the presence of non-factorizable effects such as final state interactions, or some corrections to factorization.

\section{Penguin-Diagram Mediated Decays}
The $b\to s$ transition is a penguin-diagram mediated FCNC process.
Since it is forbidden at tree level in the SM, its small amplitude
make it sensitive to effects caused by exchange of non-SM
particles in the penguin loop.

The CLEO group reported the first observation of the $\bbar\to X_s \gamma$ radiative penguin decay.\cite{sgamma-CLEO}
Here, $\bbar$ represents $\bzb$ or $\bm$ mesons and $X_s$ represents a hadron system that includes a $s$ quark.
The measured branching fraction for this process provides the most stringent indirect limit on the charged Higgs mass range\cite{higgs-mass} and a constraint on the magnitude of the effective Wilson coefficient of the electromagnetic penguin operator $|C_7^\mathrm{eff}|$.
The electromagnetic penguin decays $\bbar\to X_s \ell^+\ell^-$ are essential to further constrain the Wilson coefficients.
The magnitude and the phase of the coefficients $C_7^\mathrm{eff},\;C_9^\mathrm{eff},\;C_{10}$ can be determined by measuring the dilepton invariant mass distributions and forward-backward charge asymmetry of the dilepton and the $\bbar\to X_s \gamma$ rate.\cite{Wison-coefficients}
These measurements are crucial to obtain definitive evidence for new physics.

\bigskip

The branching fraction for $\bbar\to X_s \gamma$ has been calculated to 10\% precision in the framework of the SM including next-to-leading order QCD corrections.\cite{sgamma-SM}
It is important to measure the branching fraction to a precision of 10\% or better to explore or limit non-SM theories.

The inclusive $\bbar\to X_s \gamma$ decay is reconstructed by combining the $X_s$ system with a photon. The $X_s$ system is formed by combining a charged kaon or $\ks$ with 0--4 pions which may include one $\piz$.
Combinatorial background is reduced by requiring $M_{X_s}<2.05$~\Gevcsq.
The SFW variable is employed to suppress continuum background.
In addition, the SFW sideband is used to model the background shape for the continuum events.
The fit to the \Mb\ distribution yields $107\pm17$ events in a 5.8~\Fb\ data sample, corresponding to a branching fraction of
\begin{eqnarray}
\BF(\bbar&\to& X_s \gamma)=(3.36\pm0.53\pm0.42^{+0.50}_{-0.54})\times10^{-4}.
\end{eqnarray}
The third error is due to extrapolation of the $M_{X_s}$ spectrum from the region $M_{X_s}<2.05$~\Gevcsq.
Our result is consistent with the SM prediction\cite{sgamma-SM} of $(3.28\pm0.33)\times10^{-4}.$

In order to reduce the systematic error due to the $M_{X_s}$
spectrum, we need to understand the 
resonant structure of the ${X_s}$ system 
beyond the $\bbar\to\kstzb\gamma$ decay which has been well measured.
The experimental search for higher kaonic resonances yielded only an indication of $\bbar\to K_2^*(1430)\gamma$ so far.
In this analysis, we study the kaonic resonances which decay into $\km\pip$, $\ks\pim$, $\km\piz$, and $\km\pip\pim$ modes.

In the $\bzb\to\km\pip\gamma$ mode where the $\km\pip$ mass is
required to be consistent with $\kbar{}_2^{*0}(1430)$, a fit to the \Mb\ distribution yields $29.1\pm6.7$ events in a 21.3~\Fb\ data sample.
The helicity angle distribution is analyzed to distinguish the $\kbar{}_2^{*0}(1430)$ signal from $\kbar{}^{*0}(1410)$ and non-resonant modes.
A fit to the helicity angle distribution yields $20.1\pm10.5$ events for the $\kbar{}_2^{*0}(1430)$ component, which leads to a branching fraction of
\begin{eqnarray}
\BF(\bzb&\to&\kbar{}_2^{*0}(1430)\gamma)=(1.26\pm0.66\pm0.10)\times10^{-5}.
\end{eqnarray}

\begin{figure}
{\footnotesize \vspace*{0.05cm}\hspace*{4.4cm}$\bm$$\to$$\kstzb\pim\gamma$\hspace*{2.75cm}$\bm$$\to$$\km\rho^0\gamma$}
\vspace*{-0.90cm}
\center{\resizebox{0.60\textwidth}{!}{\includegraphics{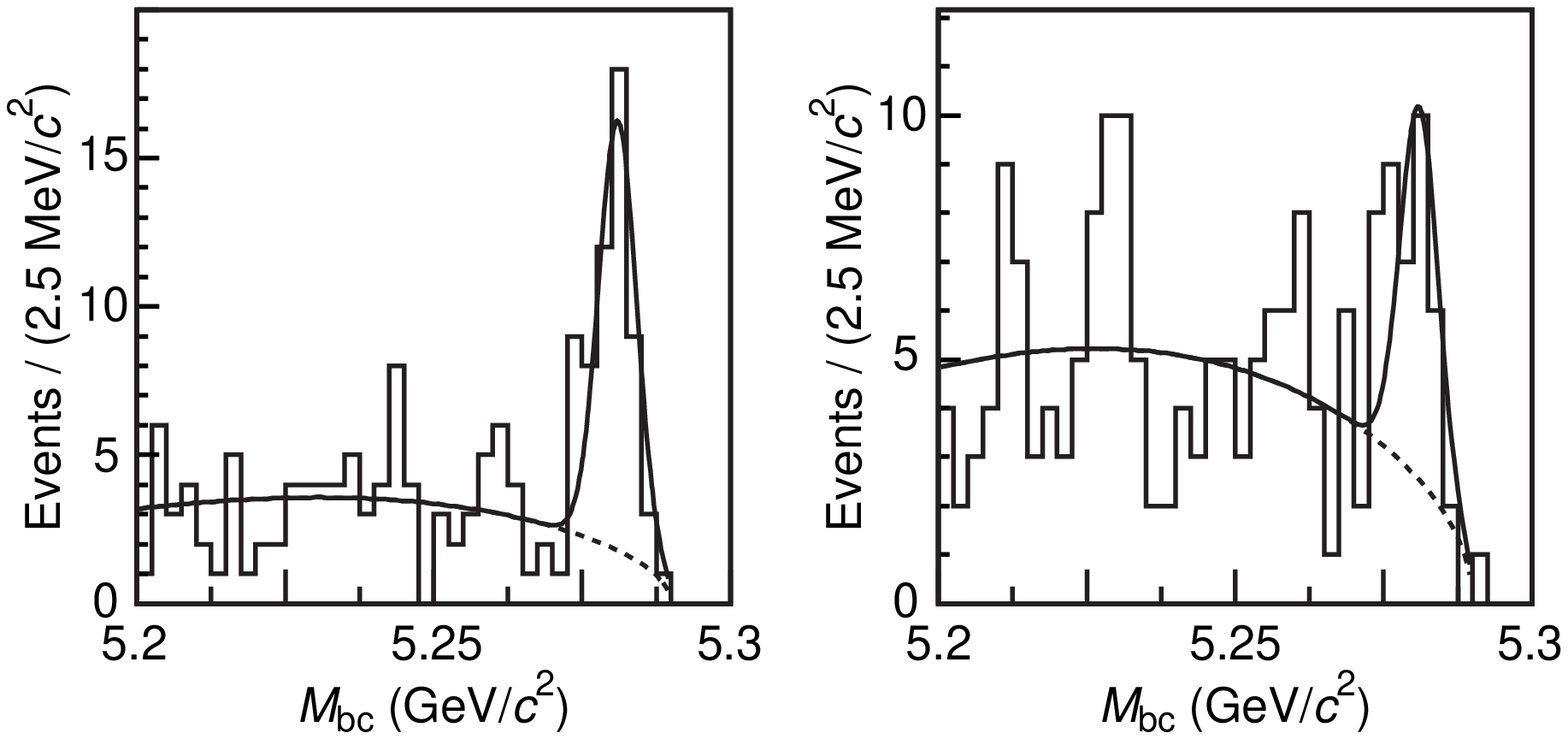}}}
\caption{\Mb\ distributions for $\bm\to\kstzb\pim\gamma$ and $\bm\to\km\rho^0\gamma$ modes.
The solid lines show the fit functions, and the dashed lines show the background.}
\label{fig:Kxgamma}
\end{figure}

We reconstruct $\bm\to\km\pip\pim\gamma$ decays in
$\bm\to\kstzb\pim\gamma$ and $\bm\to\km\rho^0\gamma$ modes where
$\km\pip$ or $\pip\pim$ mass is 
required to be consistent with $\kstzb$ or $\rho^0$.
$M_{\km\pip\pim}$ is required to be less than 2.0~\Gevcsq\ to reduce combinatorial backgrounds.
Fits to the \Mb\ distributions yield $46.4\pm7.3$ events in the $\bm\to\kstzb\pim\gamma$ mode and $24.5\pm6.4$ events in the $\bm\to\km\rho^0\gamma$ modes.
Figure~\ref{fig:Kxgamma} shows the \Mb\ distributions for $\bm\to\kstzb\pim\gamma$ and $\bm\to\km\rho^0\gamma$ modes with the fit results.
After subtracting the backgrounds such as $\bm\to\km\pip\pim\gamma$ non-resonant decays, feed-across and other $\bbar\to X_s \gamma$ decays, we obtain signal yields of $39.7\pm7.4$ events in the $\bm\to\kstzb\pim\gamma$ mode and $22.2\pm6.5$ events in the $\bm\to\km\rho^0\gamma$ modes, which gives branching fractions of
\begin{eqnarray}
\BF(\bm\to\kstzb\pim\gamma;M_{\kstzb\pim}<2.0~{\rm GeV}/c^2)&=&(5.6\pm1.1\pm0.9)\times 10^{-5},\nonumber\\
\BF(\bm\to\km\rho^0\gamma;M_{\km\rho^0}<2.0~{\rm GeV}/c^2)&=&(6.5\pm1.7^{+1.1}_{-1.2}\times 10^{-5}.
\end{eqnarray}
The sum of measured exclusive modes accounts for about a half of the total $\bbar\to X_s \gamma$ branching fraction.

\bigskip

A great deal of attention has been paid to the $\bbar\to X_s\ell^+\ell^-$ decay since it provides information necessary to determine the Wilson coefficients, $C_7^\mathrm{eff},\;C_9^\mathrm{eff}$, and $C_{10}$.
Some non-SM models such as SUSY gives significantly different values for these coefficients from the SM prediction due to non-SM particle contributions to the loop in the penguin diagram.
However, no experimental evidence has been observed for the $\bbar\to X_s \ell^+\ell^-$ decays.

We have searched for $\bbar\to X_s\ell^+\ell^-$ decay using both
an exclusive and inclusive approach.
The background from $\bbar\to\psi^{(')}K,\;\psi^{(')}\to\ell^+\ell^-$ is rejected by requiring the dilepton invariant mass to be outside of $\psi^{(')}$ mass region.
We have a wider veto region for the electron mode because the electron tends to lose its energy due to initial state radiation or bremsstrahlung.
Several event shape variables are combined into a Fisher discriminant to suppress continuum background.
The missing energy of the event is used to suppress the major background from semileptonic $B$ decays.
The Fisher discriminant variable and the missing energy are combined with kinematic variables into likelihood ratios to maximize the background rejection capability.
We reject 85\% of continuum background and 45--55\% of $\bbbar$ background while retaining 70--75\% of the signal.

In the inclusive analysis, the signal yield is obtained from a unbinned ML fit to the \Mb\ distribution with the sum of a Gaussian function for the signal and an ARGUS function for the background.
The background shape is determined by the fit while the signal shape is calibrated using the $\bbar\to J/\psi\kbar$ sample in data.
We find $11.4^{+5.1}_{-4.8}$ events with a statistical significance
of $2.7\sigma$ for the $\mu^+\mu^-$ mode in a 29.5~\Fb\ data sample.

In the exclusive analysis, the background shape is determined using a MC sample corresponding to 400~\Fb\ due to lack of statistics in the data sample.
We also take into account the background due to lepton misidentification where $\bbar\to\kbar{}^{(*)} h^+h^-$ decay is misidentified as $\bbar\to\kbar{}^{(*)}\ell^+\ell^-$ where $\kbar{}^{(*)}$ represents $\km$, $\kzb$, $\kstm(892)$ and $\kstzb(892)$.
\begin{figure}
{\footnotesize 
\vspace*{0.15cm}
\hspace*{4.45cm}$\bbar\to\kbar\mu^+\mu-$
\hspace*{2.1cm}$\bbar\to\kbar\mu^+\mu-$ MC BG\vspace*{1.55cm}\\
\hspace*{4.45cm}$\bbar\to\kbar e ^+ e^-$
\hspace*{2.2cm}$\bbar\to\kbar e ^+ e^-$ MC BG\vspace*{1.55cm}\\
\hspace*{4.45cm}$\bbar\to\kbar\ell^+\ell^-$
\hspace*{2.2cm}$\bbar\to\kbar\ell^+\ell^-$ MC BG
}
\vspace*{-5.25cm}
\center{\resizebox{0.60\textwidth}{!}{\includegraphics{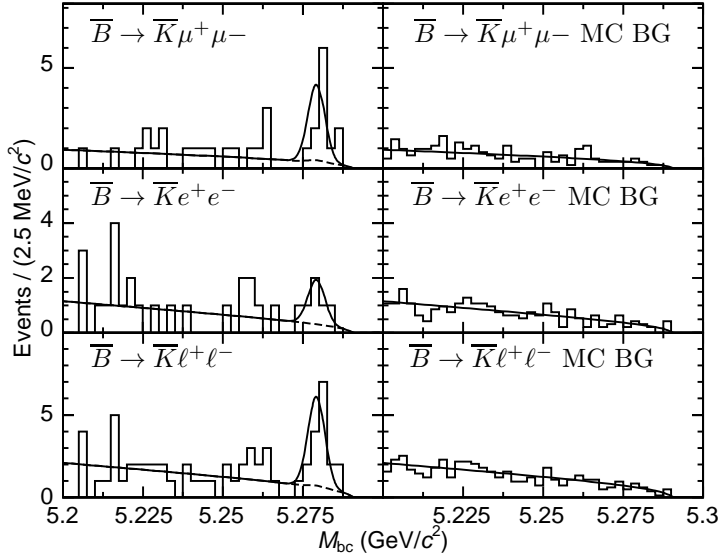}}}
\caption{\Mb\ distributions for $\bbar\to\kbar\ell^+\ell^-$ modes.
The solid lines show the fit functions, and the dashed lines show the background in the left column.
The solid lines show a fit to the ARGUS function in the right column.}
\label{fig:mb-Kll}
\end{figure}
This background is included as a small additional Gaussian background function in the fit.
We find $9.5^{+3.8}_{-3.1}$ events in the $\kbar \mu^+\mu^-$ mode, $4.1^{+2.7}_{-2.1}$ events in the $\kbar e^+e^-$ mode and $6.3^{+3.7}_{-3.0}$ events in the $\kstb e^+e^-$ mode with statistical significances of $4.7\sigma$, $2.5\sigma$ and $2.5\sigma$, respectively.
We do not observe a significant signal in the $\kstb \mu^+\mu^-$ mode.
The probability of an upward fluctuation of the background to the
observed number of events in the $\kbar \mu^+\mu^-$ mode is
$5.5\times10^{-6}$, which corresponds to $4.4\sigma$ for a 
Gaussian probability distribution.
When we combine the $\kbar \mu^+\mu^-$ and $\kbar e^+e^-$ modes, we observe $13^{+4.5}_{-3.8}$ events with a statistical significance of $5.5\sigma$ further establishing the observation of the $\bbar\to\kbar\ell^+\ell^-$ decay.
Figure~\ref{fig:mb-Kll} shows the \Mb\ distributions for the $\bbar\to\kbar\ell^+\ell^-$ modes with the fit results.
The left column shows the data and the right column shows the background distributions in the MC sample.
The branching fraction is measured to be
\begin{eqnarray}
\BF(\bbar&\to&\kbar\ell^+\ell^-)=(0.75^{+0.25}_{-0.21}\pm0.09)\times10^{-6}.
\end{eqnarray}
This result is consistent with some of the SM
predictions\cite{Kll-SM} which cover the range
(0.42--0.57)$\times10^{-6}$, although one model\cite{Kll-Greub} 
gives a slightly lower value of $(0.33\pm0.07)\times10^{-6}$.

\section{Conclusions}
We have established the existence of the
electroweak penguin-mediated decay $\bbar\to\kbar\ell^+\ell^-$, which opens a new window to search for physics beyond the SM.
We have also made the following new observations.
\begin{itemize}
\item Color-suppressed $B$ decays, $\bzb\to\dz\piz$, $\dz\eta$, $\dz\omega$.
\item Three-boy charmless $B$ decays, $\bzb\to\km h^+h^-$, in which $\bm\to\chi_{c0}\km$ decay is observed. 
The large branching fractions for $\chi_{c0}\km$ decay and the color-suppressed decays pose challenges to factorization models.
\item $\bm\to\dcp\km$ decay. We have made an asymmetry measurement, which is the first step toward a $\phi_3$ measurement.
\item $\bzb\to D^{*\pm}D^{\mp}$ decay with full and partial reconstruction technique.
This is another promising mode that can be used to measure $\phi_1$.
\item $\bm\to\kstzb\pim\gamma$ and $\bm\to\km\rho^0\gamma$. The sum
of the exclusive branching fractions accounts for about 50\% of the inclusive branching fraction.
\end{itemize}
 
In addition to the above new observations, we have observed and measured branching fractions for $\bbar\to\phi\kbar$, $\bzb\to\dstp\dstm$ and $\bbar\to\pi\pi$ decays.
These modes will provide measurements of $\phi_1$ and $\phi_2$ in
the near future.
We have also confirmed the large branching fractions for $\bbar\to\eta'\kbar$ and $\bbar\to\eta\kstb$ decays, which is another theoretical challenge.

We have made precise measurements of $|\vcb|$ via inclusive semileptonic $B$ decay and $\bzdstlnu$ decay.
We have measured a branching fraction for $\bzb\to\pip\ell^-\nub$ decay, which will lead to a $|\vub|$ measurement.
A precise measurement of  $|\vub|$ is the key to evaluate the
consistency of the Standard Model 
with the large $\sin2\phi_1$ value measured by the Belle experiment.

In conclusion, we have demonstrated that we are ready to take the
next step. We can measure all the $CP$ angles in many different modes in
addition to measuring $\sin2\phi_1$ via the $\bz\to\psi\kz$ mode,
and we have many effective tools to probe new physics 
beyond the Standard Model.

\section*{Acknowledgments}
We wish to thank the KEKB accelerator group for the excellent operation of the KEKB collider.


\begin{thebibliography}{99}
\bibitem{charge-conjugate}
Charge-conjugate modes are implied throughout this paper.
\end{thebibliography}
\end{document}